\documentclass[%
superscriptaddress,
preprint,
 amsmath,amssymb,
 aps,
 prd,
floatfix,nofootinbib
]{revtex4-1}

\usepackage{graphicx}
\usepackage{dcolumn}
\usepackage{bm}
\usepackage[normalem]{ulem}
\usepackage{tikz}
\usetikzlibrary{positioning}

\usepackage{braket}
\usepackage{multirow}
\usepackage{dsfont}
\usepackage{mathtools}
\usepackage{xcolor}
\usepackage[normalem]{ulem}

\usepackage{cancel}
\usepackage{epsfig,dsfont}
\usepackage{subfigure}
\usepackage{bm}
\usepackage{amssymb}
\usepackage{mathrsfs}
\usepackage{amsmath}
\usepackage{enumitem}
\usepackage{dcolumn}

\newcommand{\bA}{\boldsymbol{A}}
\newcommand{\bt}{\boldsymbol{t}}

\newcommand{\btau}{\boldsymbol{\tau}}
\renewcommand{\vec}[1]{\mbox{\boldmath$#1$\unboldmath}}

\newcommand{\I}{\mathrm{i}}
\begin{document}

\preprint{OU-HET-1025}

\title{Chiral-spin symmetry of the meson spectral function above $T_c$ }

\author{\vspace*{4mm}C.~Rohrhofer}
\affiliation{Department of Physics, Osaka University, Toyonaka 560-0043, Japan}
\affiliation{Institute of Physics, University of Graz, 8010 Graz, Austria}
\author{Y.~Aoki}
\affiliation{RIKEN Center for Computational Science, Kobe 650-0047, Japan}
\author{L.Ya.~Glozman}
\affiliation{Institute of Physics, University of Graz, 8010 Graz, Austria}
\author{S.~Hashimoto}
\affiliation{KEK Theory Center, High Energy Accelerator Research Organization (KEK), Tsukuba 305-0801, Japan}
\affiliation{School of High Energy Accelerator Science, The Graduate University for Advanced Studies (Sokendai), Tsukuba 305-0801, Japan}

\date{\today}

\begin{abstract}
\vspace{4mm} Recently, via calculation of
spatial correlators of $J=0,1$ isovector operators using a chirally symmetric
Dirac operator within $N_F=2$ QCD, it has been found that QCD at temperatures 
$T_c - 3 T_c$ is approximately $SU(2)_{CS}$ and $SU(4)$ symmetric.
The latter symmetry suggests that the physical degrees of freedom are chirally
symmetric quarks bound by the chromoelectric field into color singlet objects
without chromomagnetic effects.
This regime of QCD has been referred to as a Stringy Fluid.
Here we calculate correlators for propagation in time direction at a
temperature slightly above $T_c$ and find the same approximate symmetries.
This means that the meson spectral function is chiral-spin and $SU(4)$ symmetric
in the same temperature range.
\end{abstract}
\maketitle


\section{\label{sec:intro}Introduction}
Artificial truncation of the near-zero modes of the Dirac operator at zero
temperature results in the emergence of a large degeneracy in the hadron
spectrum, larger than implied by the chiral symmetry of the QCD Lagrangian
\cite{D1,D2,D3,D4}.
A symmetry group of this degeneracy, the chiral-spin $SU(2)_{CS}$  group and
its flavor extension $SU(2N_F)$, contains chiral symmetries as subgroups
\cite{G1,GP}.
These symmetries are not symmetries of the Dirac Lagrangian.
However they are symmetries of the electric interaction in a given reference
frame, while the magnetic interaction as well as the quark kinetic term break
them.
Consequently these symmetries allow us to separate the electric and magnetic
interactions in a given frame.
The emergence of the $SU(2)_{CS}$ and $SU(2N_F)$ symmetries in the hadron
spectrum upon the low mode truncation means that while the confining
chromoelectric interaction is distributed among all modes of the Dirac
operator, the chromomagnetic interaction contributes only (or at least
predominantly) to the near-zero modes.
Some unknown microscopic dynamics should be responsible for this phenomenon.

At high temperatures, above the pseudocritical temperature $T_c$, chiral
symmetry is restored due to the near-zero modes of the Dirac operator being
naturally suppressed by temperature effects~\cite{Chiu:2013wwa,Buchoff:2013nra,Tomiya:2016jwr,Suzuki:2019vzy}.
Then one could expect a natural emergence of the $SU(2)_{CS}$ and $SU(2N_F)$
symmetries in QCD above  $T_c$  \cite{G}.

In \cite{Rohrhofer:2017grg, Rohrhofer:2019qwq} we have studied a complete
set of  $J=0$  and  $J=1$  isovector correlation functions in $z$-direction
for a system with  $N_F=2$  dynamical quarks in simulations with the chirally
symmetric domain wall Dirac operator at temperatures up to  $5.5 T_c$.
Similar ensembles have been used previously for the study of the $U(1)_A$
restoration in $t$-correlators and via the Dirac eigenvalue decomposition of
correlators \cite{Cossu:2015kfa,Tomiya:2016jwr}.
We have observed emergence of approximate $SU(2)_{CS}$ and $SU(4)$ symmetries
in the spatial correlators in the temperature range  $T_c - 3 T_c$.
These symmetries of spatial correlators reflect symmetries of the QCD action since
correlation functions are driven only by the action of the theory.
Observation of approximate  $SU(2)_{CS}$  and  $SU(4)$  symmetries at
  $T_c - 3 T_c$  suggests that the physical degrees of freedom in this
temperature range are chirally symmetric quarks bound by the chromoelectric
field into color-singlet compounds without chromomagnetic effects.
Such a system is reminiscent of a ``string'', that is why the corresponding
regime of QCD at  $T_c - 3 T_c$  is referred to as a Stringy Fluid.
The chemical potential term in the QCD action has precisely the same
symmetries \cite{G2}, so one can expect that the symmetries observed in the
lattice calculations at zero chemical potential will persist at $\mu > 0$ as
well.

At $T>0$ correlators in time and spatial directions
have different physical content; temporal correlators
are connected
to the spectral density in Minkowski space via an integral
transformation.
Observation of the   $SU(2)_{CS}$  and  $SU(4)$  symmetries in $t$-correlators
would imply that the spectra of the corresponding color-singlet states in
Minkowski space have the same symmetry.
The symmetries of the $z$-correlators do suggest the same symmetries in the
spectra, albeit indirectly. 
A direct observation of these symmetries in $t$-correlators in practice is
a priori not obvious since on the lattice one has only a few lattice sites
along the time direction at high T and large discretization errors as well
as a small evolution time can easily spoil the real picture. 
Here we use  $N_t=12$  ensembles at  $T = 1.2 T_c$  and observe
clear   
 $SU(2)_{CS}$  and  $SU(4)$  symmetries in $t$-correlators.
This implies that the corresponding spectral functions in Minkowski space are
also  $SU(2)_{CS}$  and  $SU(4)$  symmetric.

\section{Chiral-spin symmetry }

The  $SU(2)_{CS}$ chiral-spin transformations for quarks are defined by 
\cite{G1,GP}

\begin{equation}
\psi(x) \; \rightarrow \; \exp\left(\frac{i}{2}\vec\Sigma \, \vec\epsilon\right)\psi(x) \; , \quad \;
\bar{\psi}(x) \; \rightarrow \; \bar{\psi}(x) \gamma_4 \exp\left(-\frac{i}{2}\vec\Sigma \, \vec\epsilon\right) \gamma_4 \; ,
\label{equ:su2cstrafos}
\end{equation}
where  $\vec\epsilon \in \mathds{R}^3$  are the rotation parameters.
For the generators  $\vec\Sigma$,
\begin{equation}
  { \bf \Sigma} = \{\gamma_k,-\I \gamma_5\gamma_k,\gamma_5\},
\label{eq:su2cs_}
\end{equation}
one has 
four different choices $\vec\Sigma = \vec\Sigma_k$ with $k = 1,2,3,4$.
Here $\gamma_k$, $k=1,2,3,4$, are hermitian Euclidean gamma-matrices, obeying
the anticommutation relations
\begin{equation}
\gamma_i\gamma_j + \gamma_j \gamma_i =
2\delta_{ij}; \qquad \gamma_5 = \gamma_1\gamma_2\gamma_3\gamma_4\,.
\label{eq:diracalgebra}
\end{equation}
The $\mathfrak{su}(2)$ algebra
\begin{equation}
[\Sigma^a,\Sigma^b]=2\I\epsilon^{abc}\Sigma^c
\end{equation}
is satisfied for any $k$. 

The choice of $k$ is fixed by the requirement that the $SU(2)_{CS}$
transformation does not mix operators with different spin, i.e.,
respects the rotational $O(3)$ symmetry in Minkowski space.
For propagators in  time direction, defined below, this implies $k=4$.

$U(1)_A$ is a subgroup of $SU(2)_{CS}$.
The  $SU(2)_{CS}$ transformations mix the left- and right-handed fermions
and different representations of the Lorentz group. 

The direct product  $SU(2)_{CS} \times SU(N_F)$  can be embedded into a
$SU(2N_F)$ group.
The chiral symmetry group of QCD,  $SU(N_F)_L \times SU(N_F)_R \times U(1)_A$,
is a subgroup of $SU(2N_F)$.

The  $SU(2)_{CS}$  and  $SU(2N_F)$  groups are not symmetries
of the Dirac equation as well of the QCD Lagrangian as a whole.
In a given reference frame the quark-gluon interaction Lagrangian in
Minkowski space can be split into temporal and spatial parts:
\begin{equation}
\overline{\psi}   \gamma^{\mu} D_{\mu} \psi = \overline{\psi}   \gamma^0 D_0  \psi 
  + \overline{\psi}   \gamma^i D_i  \psi .
\label{cl}
\end{equation}
Here $D_{\mu}$ is a covariant derivative that includes
interaction of the quark field $\psi$ with the  gluon field $\bA_\mu$,
\begin{equation}
D_{\mu}\psi =( \partial_\mu - ig \frac{\bt \cdot \bA_\mu}{2})\psi.
\end{equation}
The temporal term includes an interaction of the color-octet charge density
\begin{equation}
\bar \psi (x)  \gamma^0  \frac{\bt}{2} \psi(x) = \psi (x)^\dagger  \frac{\bt}{2} \psi(x)
\label{den}
\end{equation}
with the electric part of the gluonic gauge field. 
It is invariant under any unitary transformation acting in the Dirac and/or
flavor spaces.
In particular it is a singlet under  $SU(2)_{CS}$  and  $SU(2N_F)$ groups. 
The spatial part consists of a quark kinetic term and interaction with the
magnetic part of the gauge field.
It breaks  $SU(2)_{CS}$  and  $SU(2N_F)$.
We conclude that interaction of electric and magnetic components of the gauge
field with fermions can be distinguished by symmetry. 

In order to discuss the notions ``electric'' and ``magnetic'' one needs to
fix a reference frame.
The invariant mass of the hadron is the rest frame energy.
Consequently, to discuss physics of hadron mass generation it is natural to
use the hadron rest frame.

The  spectral density $\rho(\omega)$ is an integral transform 
\begin{equation}
  C_\Gamma(t) = \int d\omega \frac{\cosh(\omega(t - \frac{1}{2T}))}{\sinh(\omega \frac{1}{2T})} \rho_\Gamma(\omega)
\end{equation}
of the rest frame
$t$-direction Euclidean correlator
\begin{equation}
C_\Gamma(t) = \sum\limits_{x, y, z}
\braket{\mathcal{O}_\Gamma(x,y,z,t)
\mathcal{O}_\Gamma(\mathbf{0},0)^\dagger},
\label{eq:momentumprojection}
\end{equation}
where $\mathcal{O}_\Gamma(x,y,z,t)$ is an operator that creates a
quark-antiquark pair for mesons with fixed quantum numbers. 

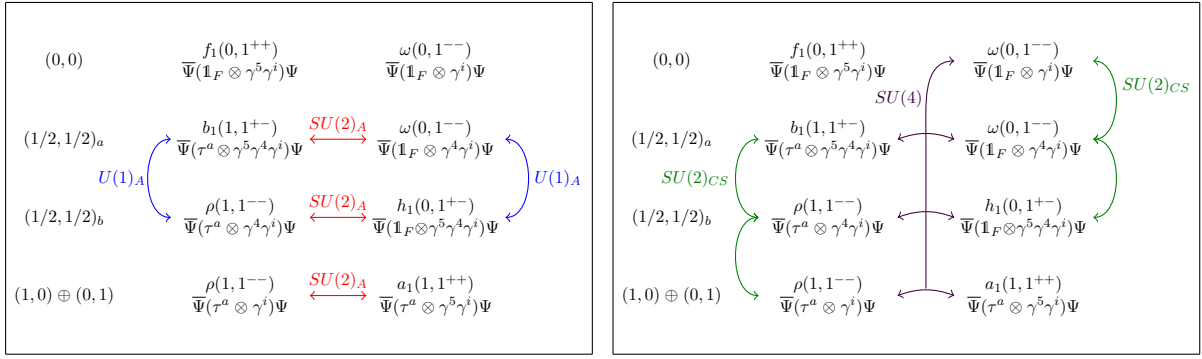
\begin{figure}
  \centering
  \begin{tikzpicture}[scale=0.52,every node/.style={scale=0.60,align=center,text width=7.3em}]

    \node (f)   at (6,8)  {$f_1(0,1^{++})$\\[-0.7em] $\overline{\Psi}(\mathds{1}_F \otimes \gamma^5\gamma^i)\Psi$};
    \node (w1)  at (11,8)  {$\omega(0,1^{--})$\\[-0.7em] $\overline{\Psi}(\mathds{1}_F \otimes \gamma^i)\Psi$};
    \node (b)   at (6,6)  {$b_1(1,1^{+-})$\\[-0.7em] $\overline{\Psi}(\tau^a \otimes \gamma^5\gamma^4 \gamma^i)\Psi$};
    \node (w2)  at (11,6)  {$\omega(0,1^{--})$\\[-0.7em] $\overline{\Psi}(\mathds{1}_F \otimes \gamma^4\gamma^i)\Psi$};

    \node (r1)  at (6,4)  {$\rho (1,1^{--})$\\[-0.7em] $\overline{\Psi}(\tau^a \otimes \gamma^4\gamma^i)\Psi$};
    \node (h)   at (11,4)  {$h_1(0,1^{+-})$\\[-0.7em] $\overline{\Psi}(\mathds{1}_F \otimes \gamma^5\gamma^4\gamma^i)\Psi$};
    \node (r2)  at (6,2)  {$\rho(1,1^{--})$\\[-0.7em] $\overline{\Psi}(\tau^a \otimes \gamma^i)\Psi$};
    \node (a)   at (11,2)  {$a_1(1,1^{++})$\\[-0.7em] $\overline{\Psi}(\tau^a \otimes \gamma^5\gamma^i)\Psi$};

    \node (cp1) at (1.5,8) {$(0,0)$};
    \node (cp2) at (1.5,6) {$(1/2,1/2)_a$};
    \node (cp3) at (1.5,4) {$(1/2,1/2)_b$};
    \node (cp4) at (1.5,2) {$(1,0)\oplus (0,1)$};

    \draw (0,0.5) -- (15,0.5);
    \draw (15,0.5) -- (15,9.5);
    \draw (15,9.5) -- (0,9.5);
    \draw (0,9.5) -- (0,0.5);

		\path[<->,blue] (b.west) edge [out=180,in=180] node[anchor=east,text width=2.5em] {$U(1)_A$} (r1.west); 
		\path[<->,blue] (w2.east) edge [out=0,in=0] node[anchor=right,right,text width=1em] {$U(1)_A$} (h.east); 

		\path[<->,red] (b.east) edge [out=0,in=180] node[anchor=center,above] {$SU(2)_A$} (w2.west); 
		\path[<->,red] (r1.east) edge [out=0,in=180] node[anchor=center,above] {$SU(2)_A$} (h.west); 
		\path[<->,red] (r2.east) edge [out=0,in=180] node[anchor=center,above] {$SU(2)_A$} (a.west); 

  \end{tikzpicture}
  \begin{tikzpicture}[scale=0.52,every node/.style={scale=0.60,align=center,text width=7.3em}]

    \node (f)   at (5.5,8)  {$f_1(0,1^{++})$\\[-0.7em] $\overline{\Psi}(\mathds{1}_F \otimes \gamma^5\gamma^i)\Psi$};
    \node (w1)  at (10.5,8)  {$\omega(0,1^{--})$\\[-0.7em] $\overline{\Psi}(\mathds{1}_F \otimes \gamma^i)\Psi$};
    \node (b)   at (5.5,6)  {$b_1(1,1^{+-})$\\[-0.7em] $\overline{\Psi}(\tau^a \otimes \gamma^5\gamma^4 \gamma^i)\Psi$};
    \node (w2)  at (10.5,6)  {$\omega(0,1^{--})$\\[-0.7em] $\overline{\Psi}(\mathds{1}_F \otimes \gamma^4\gamma^i)\Psi$};

    \node (r1)  at (5.5,4)  {$\rho (1,1^{--})$\\[-0.7em] $\overline{\Psi}(\tau^a \otimes \gamma^4\gamma^i)\Psi$};
    \node (h)   at (10.5,4)  {$h_1(0,1^{+-})$\\[-0.7em] $\overline{\Psi}(\mathds{1}_F \otimes \gamma^5\gamma^4\gamma^i)\Psi$};
    \node (r2)  at (5.5,2)  {$\rho(1,1^{--})$\\[-0.7em] $\overline{\Psi}(\tau^a \otimes \gamma^i)\Psi$};
    \node (a)   at (10.5,2)  {$a_1(1,1^{++})$\\[-0.7em] $\overline{\Psi}(\tau^a \otimes \gamma^5\gamma^i)\Psi$};

    \node (cp1) at (1.5,8) {$(0,0)$};
    \node (cp2) at (1.5,6) {$(1/2,1/2)_a$};
    \node (cp3) at (1.5,4) {$(1/2,1/2)_b$};
    \node (cp4) at (1.5,2) {$(1,0)\oplus (0,1)$};

    \draw (0,0.5) -- (15,0.5);
    \draw (15,0.5) -- (15,9.5);
    \draw (15,9.5) -- (0,9.5);
    \draw (0,9.5) -- (0,0.5);

		\path[<->,green!50!black] (b.west) edge [out=180,in=180] node[anchor=east,text width=4em] {$SU(2)_{CS}$} (r1.west); 
		\path[<->,green!50!black] (r1.west) edge [out=180,in=180] node[] {} (r2.west); 
		\path[<->,green!50!black] (w2.east) edge [out=0,in=0] node[] {} (h.east); 
		\path[<->,green!50!black] (w2.east) edge [out=0,in=0] node[anchor=south west,text width=1em] {$SU(2)_{CS}$} (w1.east); 

    \path[<-,violet!50!black] (w1.west) edge [out=180,in=90] node[left,text width=2.7em] {\hspace{3em}$SU(4)$} (8.0,6);
		\path[-,violet!50!black] (8.0,6) edge [out=270,in=90] node[] {} (8.0,2.18);
		\path[<->,violet!50!black] (b.east) edge [out=20,in=160] node[] {} (w2.west);
		\path[<->,violet!50!black] (r1.east) edge [out=20,in=160] node[] {} (h.west);
		\path[<->,violet!50!black] (r2.east) edge [out=20,in=160] node[] {} (a.west);

  \end{tikzpicture}
  \caption{Transformations between interpolating vector operators, $i=1,2,3$.
  The left columns indicate the chiral representation for each operator.
  Red and blue arrows connect operators that transform into
  each other under $SU(2)_L \times SU(2)_R$ and  $U(1)_A$, respectively.
  Green arrows connect operators that form triplets of $SU(2)_{CS}, k=4$.
  The $f_1$ and $a_1$ operators are the $SU(2)_{CS}, k=4$ -- singlets.
  Purple arrows show the 15-plet of $SU(4)$. The $f_1$ operator is
  a $SU(4)$-singlet.}
  \label{transformations}
\end{figure}

Transformation properties of the local  $J=1$  quark-antiquark bilinears 
$\mathcal{O}_\Gamma(x,y,z,t)$
with respect to $SU(2)_L \times SU(2)_R$ and  $U(1)_A$ are given on the
left side of Fig.~\ref{transformations} and those with respect to
 $SU(2)_{CS}, k=4$  and  $SU(4)$  on the right side of
Fig.~\ref{transformations} \cite{GP}.
Emergence of the respective symmetries is signalled by the degeneracy of
the correlators (\ref{eq:momentumprojection}) calculated with operators
that are connected by the corresponding transformations.
%

\section{Methodology}

The lattice data presented in the next section is calculated on JLQCD
gauge configurations with $N_F=2$ fully dynamical domain wall fermions
(\cite{Tomiya:2016jwr,Kaneko:2013jla}).
The length of the fifth dimension for the fermions is chosen as
$L_5=16$, to ensure good chiral symmetry~\cite{Cossu:2015kfa}.

The quark propagators are computed on point sources after three steps of stout smearing.
The fermion fields obey anti-periodic boundary conditions in time direction.
For the gauge part we use the Symanzik-improved gauge action with an
inverse gauge coupling $\beta_g = 4.3$ ($a=0.075$ fm).
The time extent of the lattices is $N_t=12$, which corresponds to a temperature
of $T\simeq 220$ MeV ($\sim 1.2 T_c$).
We calculate the data on three spatial volumes, $N_s=24, 32, 48$,
with a quark mass of $m_{ud}=0.001$.
Measurements are performed on $\mathcal{O}(50)$ independent configurations.

The main observables are correlation functions of local isovector bilinears
$$
\mathcal{O}_\Gamma= \bar\psi \left(\btau/2 \otimes \Gamma\right)\psi,
$$
where the $\Gamma$ structures from Fig.~\ref{transformations} determine
the resulting quantum numbers.
To extract correlation functions of states with definite, i.e. zero, momentum,
we perform a momentum projection according to Eq.(~\ref{eq:momentumprojection}).

Finally, the data shown in the next section is rescaled to a dimensionless variable
\begin{align}
t\,T \; = \; (n_t a)/(N_t a) \; = \; n_t/N_t \; ,
\label{z_dimless}
\end{align}
where~$t$ is the measured lattice distance in time direction, $T$ the temperature,
$a$ the lattice constant, and $N_t$ the overall temporal lattice extent.
For spatial  correlators in $z$-direction the same rescaling is done
with $z=n_z a$ instead of $t$.

\section{Results}

\begin{figure}
  \centering
  \includegraphics[width=0.49\linewidth]{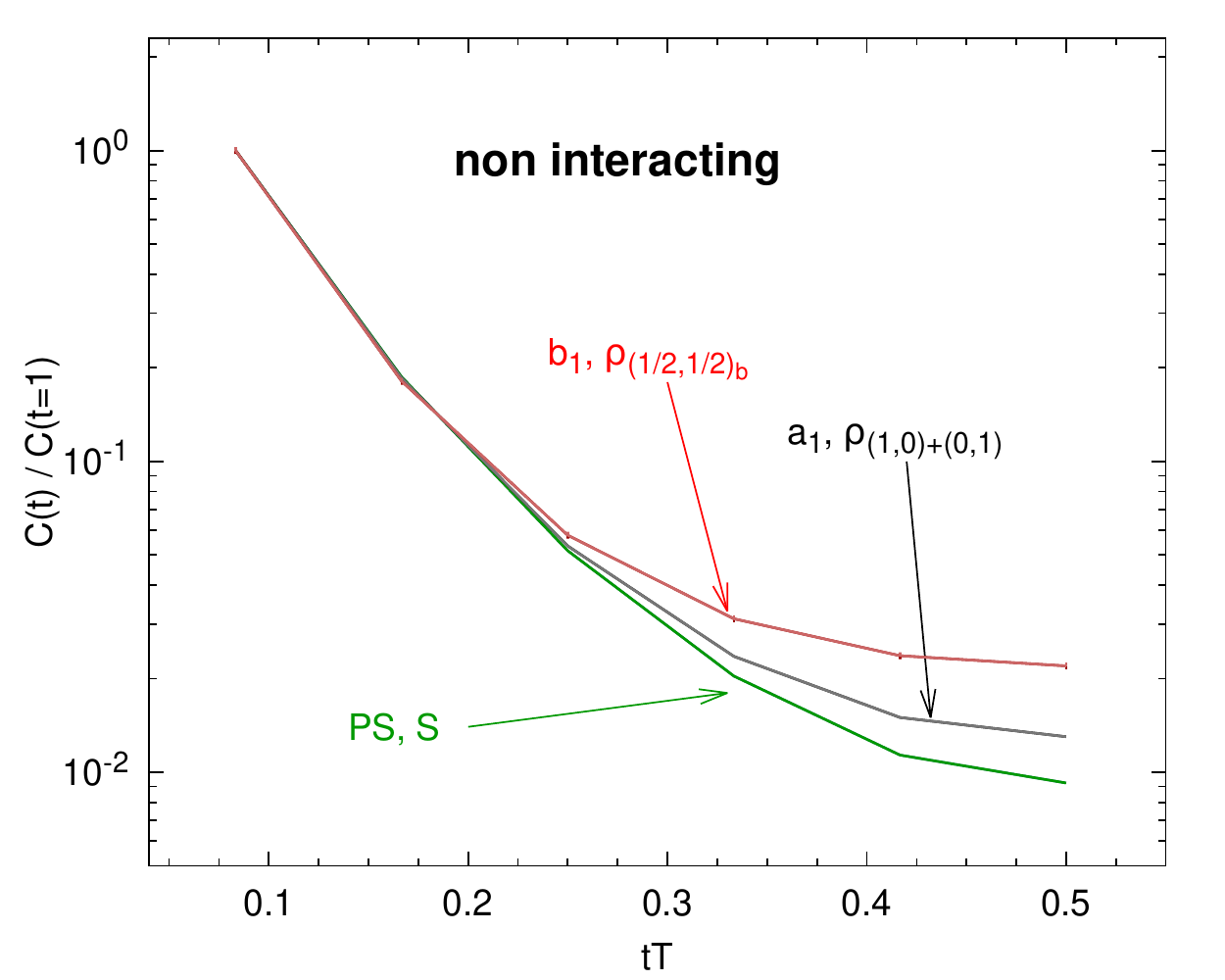}
  \includegraphics[width=0.49\linewidth]{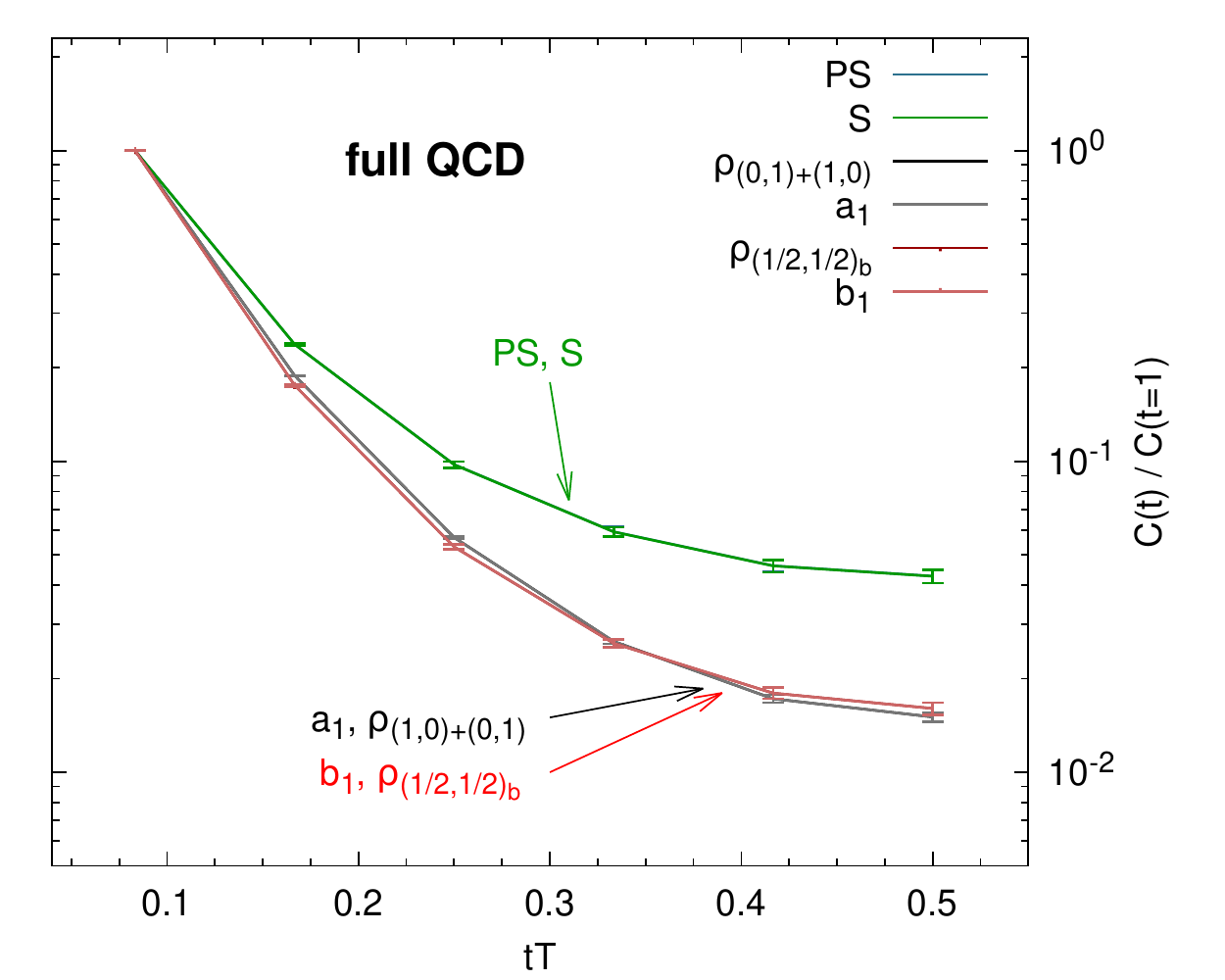}
  \caption{Temporal correlation functions for $48^3\times 12$ lattices.
          The l.h.s. shows correlators calculated with free noninteracting quarks on the
          same lattice, and features a
          symmetry pattern expected from chiral symmetry.
          The r.h.s. presents full QCD data at a temperature of $T=220$MeV
	  ($1.2 T_c$),
          which shows multiplets of all $U(1)_A$, $SU(2)_L \times SU(2)_R$,
	  $SU(2)_{CS}$ and $SU(4)$ groups.}
  \label{tcorrs}
\end{figure}

On the right side of Fig.~\ref{tcorrs} we show $t$-correlators (\ref{eq:momentumprojection}) normalized at $n_t=1$
calculated on  $48^3 \times 12$ lattices at $T=1.2 T_c$.
The results obtained on $N_s=32,24$ lattices are similar and agree within statistical
errors, they are omitted for clarity.

Specifically we calculate the correlators of $J=0$ isovector
scalar $ \bar \psi  \btau/2 \psi$ ($S$) and pseudoscalar
$ \bar \psi \gamma_5 \btau/2 \psi$ ($PS$) operators, where $\btau$ are isospin
Pauli matrices as well as correlators of isovector operators
$\{b_1, (1/2,1/2)_a\}$, $\{\rho, (1/2,1/2)_b\}$, $\{\rho, (1,0)\oplus (0,1)\}$ 
and $\{a_1, (1,0)\oplus (0,1)\}$ from Fig. \ref{transformations}.
A degeneracy of scalar and pseudoscalar correlators reflects restoration
of $U(1)_A$ symmetry, since the corresponding operators are connected
by $U(1)_A$ transformations, observed already previously in Refs. \cite{Cossu:2015kfa,Tomiya:2016jwr}.
Since the $\{b_1, (1/2,1/2)_a\}$ and $\{\rho, (1/2,1/2)_b\}$ operators are
also connected by $U(1)_A$ transformations, the degeneracy of the
corresponding correlators also signals $U(1)_A$ symmetry.
A degeneracy of $\{\rho, (1,0)\oplus (0,1)\}$ and 
$\{a_1, (1,0)\oplus (0,1)\}$ correlators evidences the restoration of
chiral $SU(2)_L \times SU(2)_R$ symmetry.

An approximate degeneracy of $\{b_1, (1/2,1/2)_a\}$, $\{\rho, (1/2,1/2)_b\}$ and $\{\rho, (1,0)\oplus (0,1)\}$ correlators signals emergence of $SU(2)_{CS}$
symmetry, since all three operators belong to the same irreducible
representation (triplet) of  $SU(2)_{CS}$.
Finally a degeneracy of
all four correlators $\{b_1, (1/2,1/2)_a\}$, $\{\rho, (1/2,1/2)_b\}$, $\{\rho, (1,0)\oplus (0,1)\}$ and $\{a_1, (1,0)\oplus (0,1)\}$
is due to the emergence of $SU(4)$ symmetry.

 On the left side of Fig.~\ref{tcorrs} we show the correlators calculated with
 free, noninteracting quarks on the same lattice with the same Dirac
 action (the gauge operator $U$ is set to 1). Dynamics
 of free quarks are governed by the Dirac equation and only chiral
 symmetries exist. Indeed, a multiplet structure in this case
 is very different as compared to the right side of Fig.~\ref{tcorrs} and only
 degeneracies due to $U(1)_A$ and $SU(2)_L \times SU(2)_R$ symmetries
 are seen in meson correlators calculated for free quarks. The pattern
 seen on the left of Fig.~\ref{tcorrs} reflects correlators at a very high temperature,
 since due to the asymptotic freedom at very high $T$ the quark-gluon
 interactions can be neglected.

While we observe practically exact chiral symmetries, the
$SU(2)_{CS}$ and $SU(4)$ symmetries are only approximate.
A degree of the symmetry breaking can be evaluated via the
parameter~$\kappa$,
\begin{equation}
  \label{def_kappa}
  \kappa = \frac{C_{\rho}^{(1,0)\oplus (0,1)} - C_{\rho}^{(1/2,1/2)}}{C_{\rho}^{(1,0)\oplus (0,1)} - C_{S}},
\end{equation}
that measures the splitting within the $SU(2)_{CS}$ multiplet
relative to the distance between different multiplets.
With this definition, good symmetry implies $|\kappa| \ll 1$.

\begin{figure}
  \centering
  \includegraphics[scale=0.7]{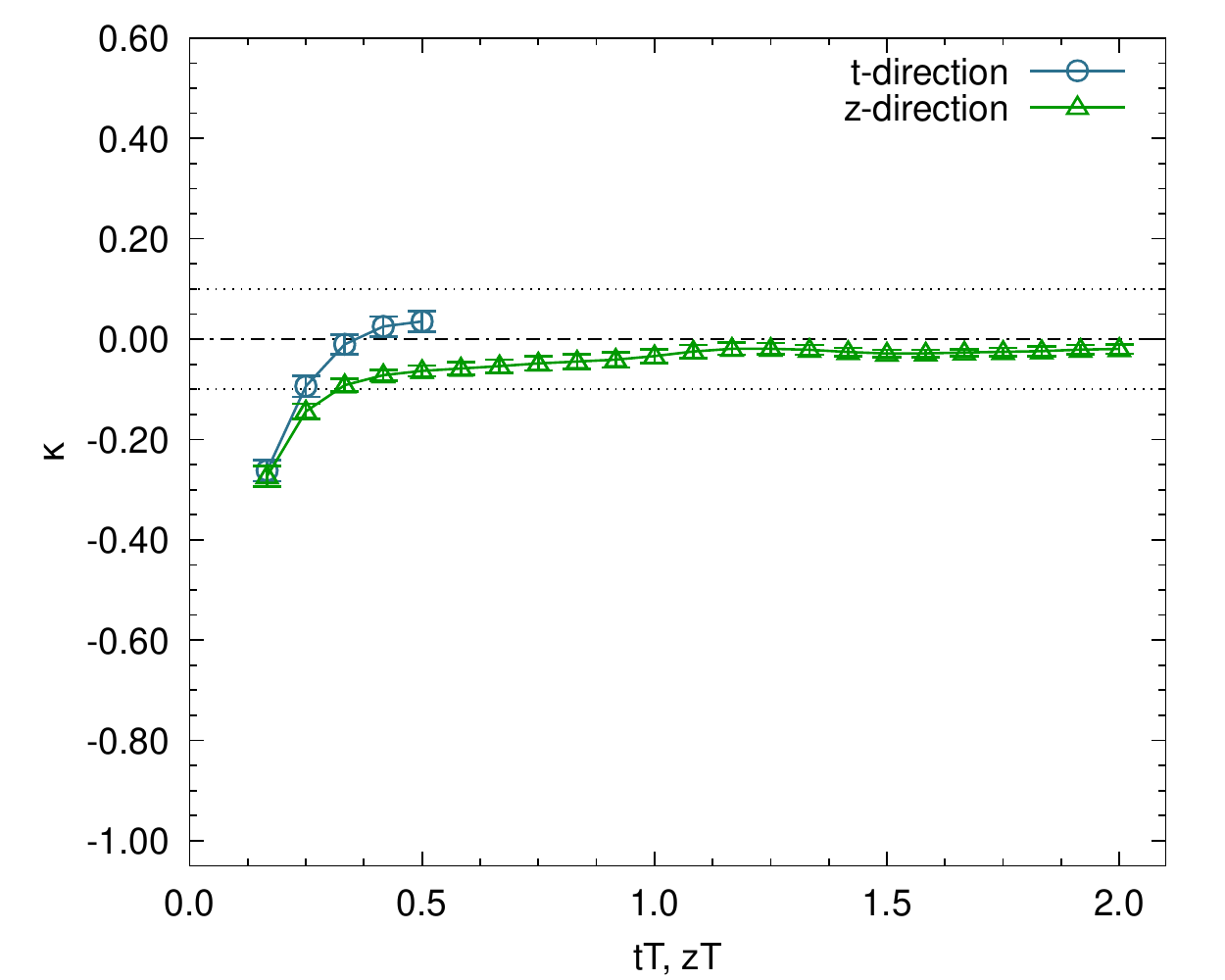}
  \caption{Kappa parameter for $48^3\times 12$ lattice at $T=220$ MeV.
           The blue circles show $\kappa$ for $t$-correlations, the
           green triangles for longer $z$-correlations.
           Both values saturate at $|\kappa| < 0.05$.}
  \label{kappa}
\end{figure}

The degree of the symmetry breaking obviously depends on the dimensionless
variable $tT$.
At $tT \sim 0.5$ the breaking is tiny, as can be seen
from Fig.~\ref{kappa}.
For the noninteracting quarks there is no $SU(2)_{CS}$ symmetry and
in infinite volume $|\kappa | \sim 1$ \cite{Rohrhofer:2019qwq}.

It is instructive to compare the scale dependence of the
symmetry breaking parameter $\kappa$ extracted from
$t$-correlators and from $z$-correlators \cite{Rohrhofer:2019qwq}
since $t$- and $z$-correlators probe QCD at different dimensionless
``distance'' $tT$, $zT$ (the time extent of the lattice is smaller
than its spatial extent).
In our finite temperature setup ($T>0$)
$t$- and $z$-correlators have different
sensitivity to electric and magnetic fields.
E.g. in the
limiting case $T \rightarrow \infty$
QCD reduces to a 3-dim magnetic theory
with a vanishing coupling constant.
Consequently at $T \sim 1.2 \; T_c$ a possible
small admixture of the magnetic field should
have a larger symmetry breaking effect in the $z$-direction
than in the $t$-direction correlator.
This is well visible in the symmetry
breaking parameter $\kappa$ at $tT=zT=0.5$.

\section{Conclusions}

We have calculated meson rest-frame correlators of $J=0$ and $J=1$ isovector
operators along the time-direction with $N_F=2$ QCD with physical masses
with the chirally symmetric domain wall Dirac operator at $T=1.2 T_c$.
We have observed a very clear emergence of approximate chiral-spin $SU(2)_{CS}$
and $SU(4)$ symmetries in these correlaros.
The $t$-correlators are connected via an integral transform with the
measurable spectral density in Minkowski space.
Approximate $SU(2)_{CS}$ and $SU(4)$ symmetries of the $t$-correlators
imply the same symmetries of spectral densities.
This result reinforces our findings in Refs.~\cite{Rohrhofer:2017grg,Rohrhofer:2019qwq}.

These symmetries are incompatible with free deconfined quarks and
suggest that the physical degrees of freedom are chirally symmetric
quarks bound into color-singlet compounds by the chromoelectric
field without chromomagnetic effects. This result
relies solely on lattice results and symmetry classification of the
QCD Lagrangian.
Such relativistic objects are reminiscent of ``strings'' since they
are purely electric and we refer to the corresponding regime of QCD as
a Stringy Fluid.

\begin{acknowledgments}

The authors thank T. Cohen, H. Fukaya, C. Gattringer, C. B. Lang and R. Pisarski
  for discussions of results.
Numerical simulations are performed on IBM System Blue Gene Solution at KEK
under a support of its Large Scale Simulation Program (No. 16/17-14)
  and
Oakforest-PACS at JCAHPC under a support of the HPCI System Research Projects (Project ID:hp170061). 
This work is supported in 
part by JSPS KAKENHI Grant Number JP26247043 and by the Post-K 
supercomputer project through the Joint Institute for Computational 
Fundamental Science (JICFuS).

\end{acknowledgments}


\end{document}